\documentclass[onecolumn,preprintnumbers,superscriptaddress]{revtex4}
\usepackage{graphicx}
\usepackage{amssymb}
\usepackage{color}
\usepackage{enumerate}

\newcommand{\be}{\begin{equation}}
\newcommand{\ee}{\end{equation}}
\newcommand{\bea}{\begin{eqnarray}}
\newcommand{\eea}{\end{eqnarray}}

\newcommand{\beaa}{\begin{equation} \begin{array}{ll}}
\newcommand{\eeaa}{\end{equation} 	\end{array} }

\begin{document}

\title{Nonlinear stability of Minkowski spacetime in Nonlocal Gravity}

\author{Fabio Briscese}\email{briscese.phys@gmail.com, briscesef@sustech.edu.cn}

\affiliation{Academy for Advanced Interdisciplinary Studies, Southern University of Science	and Technology, 1088 Xueyuan Avenue, Shenzhen 518055, P.R. China.}

\affiliation{Department of Physics, Southern University of Science	and Technology, 1088 Xueyuan Avenue, Shenzhen 518055, P.R. China.}

\affiliation{Istituto Nazionale di Alta Matematica Francesco
	Severi, Gruppo Nazionale di Fisica Matematica, Citt\`{a}
	Universitaria, P.le A. Moro 5, 00185 Rome, Italy.}

\author{Leonardo Modesto}\email{lmodesto@sustech.edu.cn}

\affiliation{Department of Physics, Southern University of Science	and Technology, 1088 Xueyuan Avenue, Shenzhen 518055, P.R. China.}

\begin{abstract}
We prove that the Minkowski spacetime is stable at nonlinear level and to all perturbative orders in the gravitational perturbation in a general class of nonlocal gravitational theories that are unitary and finite at quantum level.

%We consider a class of nonlocal gravitational models and show that, under certain conditions on the nonlocal form factor, they admit all and only the vacuum solutions of general relativity (GR). Moreover, we prove that the stability of such vacuum solutions is the same as in GR.   Surprisingly, the mentioned conditions on the class of nonlocality coincide with those required to achieve the unitarity and the super-renormalizability or the finiteness of the model.      
\end{abstract}

\maketitle

%\section{Introduction}\label{introduction}

A class of nonlocal generalizations of the Einstein-Hilbert theory for gravity has been proposed and extensively studied in the last years \cite{modesto,modestoLeslaw,universality,Buoninfante:2018lnh,shapiro3}, see \cite{Review} for review. A nonlocal gravitational model have been proposed for the first time by Krasnikov in 1988 and studied by Kuz'min in 1989 \cite{kuzmin}. %, but they were abandoned since they turn out to be non unitary.  
More recently, it has been proved that nonlocal gravity (NLG) is finite in odd dimension, while a slightly modification of the theory turns out to be finite in even dimension too \cite{modesto,modestoLeslaw,universality,Review}. The unitarity issue in nonlocal field theory has been addressed in \cite{brisceseUnitarity} where it has been proved that perturbative unitarity is preserved at any order in the loop expansion.

At classical level, all the solutions of Einstein's gravity in vacuum are solutions of NLG too \cite{yaudong}. Moreover, it has been shown that a Ricci flat spacetime is stable under linear perturbations if it is stable in Einstein's gravity  \cite{CM,CMM} (see also \cite{CN}). Therefore, based on the Wheeler-Regge result, it turns out that the Schwarzschild spacetime is stable in NLG at linear level.

In the context of Einstein's gravity, the global stability of the Minkowski spacetime has been established long time ago \cite{Christodoulou}, see also \cite{Christodoulou2}. In particular, any Strongly Asymptotically Flat (SAF) initial data set satisfying a Global Smallness Assumption (GSA) evolves in a smooth, geodesically complete and asymptotically flat solution of vacuum Einstein's equations; see the Appendix A \ref{definition SAF GSA} for a definition of the SAF condition and the GSA, and a discussion of the stability theorem of Minkowski spacetime in general relativity. Therefore, the SAF condition and the GSA  define the class of small perturbations under which the Minkowski metric is stable.  In this paper we show that this theorem is still valid in NLG. 

We consider the following minimal nonlocal action for the gravitational field, 
\be
S_{\rm g}  =  - \frac{2}{\kappa^{2}_D}  \int  d^4 x \sqrt{-g} \left[ R + G_{\mu\nu} \gamma(\Box) R^{\mu\nu}+V(\mathcal{R}) \right] \, .
\label{action gravity}
\ee
where $R$, $ R_{\mu\nu}$ and $G_{\mu\nu}$  are the Ricci scalar, Ricci curvature and the  Einstein tensor respectively. Moreover, $V(\mathcal{R})$ is a generalized potential at least cubic  in the Ricci and/or Einstein's tensor, while $\mathcal{R}$ stays for scalar, Ricci or Riemann curvatures, and derivatives thereof.
Finally, the form factor $\gamma(\Box)$ depends on the non locality scale $\ell \equiv \sqrt{\sigma}$ and is defined by 
\be\label{definition gamma}
\gamma(\Box) \equiv \frac{f(\sigma \Box) -1}{\Box} \, ,
\ee
where $f(z)$ is an entire analytic function without zeros  for finite  complex $z$, e.g. $f(z)= \exp {H(z)}$.

The equations of motion for the action (\ref{action gravity}) have been derived in \cite{eom} and read \footnote{We signal an imprecision in eq.(A.22) in \cite{eom}, which is valid only for the Minkowski metric, while (A.19-A.20) are valid in general. } 
\bea \label{einstein equation I}
%&& %\hspace{-.5cm}
E_{\mu\nu} \equiv \left(1 + \Box \, \gamma\left(\Box \right) \right) G_{\mu\nu}+\left(g_{\mu\nu} \nabla_\alpha \nabla_\beta - g_{\alpha\mu} \nabla_\beta \nabla_\nu\right)\gamma\left(\Box \right) G^{\alpha\beta}   + %\nonumber\\
%&& =  
Q_{2\mu\nu}\left({\rm Ric}\right) =8 \pi  G_{\rm N} T_{\mu\nu} \, ,
\eea
where $T_{\mu\nu} \equiv -(2/\sqrt{-g}) \delta S_m/\delta{g_{\mu\nu}}$ is the matter stress-energy tensor. Moreover, $Q_{2}({\rm Ric})$ is a sum of local and nonlocal analytic terms at least quadratic in the Ricci tensor and/or the Ricci scalar \cite{eom}, e.g. 
\be
\sigma  \left(\left(\sigma\Box\right)^n R_{\mu\alpha}\right)\left(\left(\sigma\Box\right)^m R^{\alpha}_{\,\,\,\nu}\right) \quad {\rm or} \quad \sigma^2 \left(\left(\sigma\Box\right)^n R_{\mu\alpha}\right)\left(\left(\sigma\Box\right)^m R^{\alpha}_{\,\,\,\nu}\right)(\left( \sigma \Box\right)^l R) \, , 
\ee
for integer $n, m, l$ (the label $2$ stays exactly for at least quadratic in the Ricci tensor). 
However, regardless of the explicit form of $Q_2({\rm Ric})$, in what follows it will be only relevant that $Q_2({\rm Ric})$ is at least quadratic in ${\rm Ric}$, which implies the following perturbative expansion: $Q_2\left({\rm  \, Ric}\right) = O(\epsilon^{2 n})$ if ${\rm Ric} = O(\epsilon^{n})$ for $\epsilon \ll 1$.

Since we are interested in the stability of Minkowski spacetime in vacuum, hereafter we set $T_{\mu\nu}=0$.  We start our analysis  considering small perturbations of the Minkowski metric, i.e.,
\be\label{definition h}
g_{\mu\nu} = \eta_{\mu\nu} + \epsilon \, h_{\mu\nu}, \qquad {\rm with} \qquad  |\epsilon \, h_{\mu\nu}| \ll 1
\ee
where $\eta_{\mu\nu}$ is the Minkowski metric and  $\epsilon \ll 1$ is a small dimensionless parameter.  We then proceed to determine the vacuum solutions of (\ref{einstein equation I}) perturbatively. We will  show that equations (\ref{einstein equation I}) are verified iff the Einstein's tensor $G_{\mu\nu}$ vanish at any perturbative order in $\epsilon$. Therefore, we conclude that the evolution of small perturbations of Minkowski spacetime is the same in NLG and in general relativity. This implies that the stability of Minkowski spacetime against small perturbations is the same in NLG and Einsteins gravity.

In order to obtain the perturbative expansion of the equations of motion (\ref{einstein equation I}), we start expanding in power series in the small parameter $\epsilon$ the tensors and tensor operators that appeared in (\ref{einstein equation I}), see \cite{Weinberg} for a review. We first expand $h_{\mu\nu}$ and then the Einstein's tensor $G_{\mu\nu}$, namely 
\be 
\label{definition perturbation h and G}
h_{\mu\nu} = \sum_{n=0}^\infty \epsilon^n h^{(n)}_{\mu\nu} \quad {\rm and}  \quad  G_{\mu\nu}(g_{\mu\nu}) = \sum_{n=1}^\infty \epsilon^n G^{(n)}_{\mu\nu} \, .
\ee
Notice that the leading contribution to the Einstein tensor is of order $\epsilon$ because 
$G^{(0)}_{\mu\nu} \equiv G_{\mu\nu}(\eta)=0$. Similarly,  we can expand the covariant derivative as 
\be
\label{definition perturbation nabla}
\nabla_\alpha = \sum_{n=0}^{\infty} \epsilon^n \nabla^{(n)}_\alpha = \partial_\alpha + \sum_{n=1}^{\infty} \epsilon^n \nabla^{(n)}_\alpha \, ,
\ee
where we have shown explicitly only the first term $\nabla^{(0)}_\alpha = \partial_\alpha$, since the other terms will not be needed. 
Now let us define and expand the following differential operator in (\ref{einstein equation I})
\bea\label{definition expansion A}
A_{\mu\nu\alpha\beta}\equiv \left(g_{\mu\nu} \nabla_\alpha \nabla_\beta - g_{\alpha\mu} \nabla_\beta \nabla_\nu\right)\gamma\left(\Box \right) = \sum_{n=0}^{\infty} \epsilon^n A^{(n)}_{\mu\nu\alpha\beta}
= 
\left(\eta_{\mu\nu} \partial_\alpha \partial_\beta - \eta_{\alpha\mu} \partial_\beta \partial_\nu\right)\gamma\left(\Box^{(0)} \right)
+ \sum_{n=1}^{\infty} \epsilon^n A^{(n)}_{\mu\nu\alpha\beta} \,  ,
\eea 
%where the first term in (\ref{definition expansion A}) is
%
%\bea\label{definition A0}
%A^{(0)}_{\mu\nu\alpha\beta} = \left(\eta_{\mu\nu} \partial_\alpha \partial_\beta - \eta_{\alpha\mu} \partial_\beta \partial_\nu\right)\gamma\left(\Box^{(0)} \right) \, ,
%\eea
and $\Box^{(0)}= \eta^{\rho\sigma}\partial_\rho\partial_\sigma $. We also define and expand the following differential operator again present in (\ref{einstein equation I}), namely 
\bea
\label{definition expansion f}
f(\sigma \Box) \equiv \left(1 + \Box \, \gamma\left(\Box \right) \right) = \sum_{n=0}^{\infty} \epsilon^n f^{(n)} 
= 1 + \Box^{(0)} \, \gamma\left(\Box^0 \right) + \sum_{n=1}^{\infty} \epsilon^n f^{(n)}
= f(\sigma \Box^{(0)}) + \sum_{n=1}^{\infty} \epsilon^n f^{(n)}  \, ,
\eea
%where the first term in (\ref{definition expansion f}) will be 
%
%\bea\label{definition expansion f 0}
%f^{(0)}= 1 + \Box^0 \, \gamma\left(\Box^0 \right) = f(\sigma \Box^0) .
%\eea
From the definitions and expansions (\ref{definition expansion A}) and (\ref{definition expansion f}) it follows that 
\bea 
\label{expansion A G }
%&&\hspace{-.5cm}
 A_{\mu\nu\alpha\beta} G^{\alpha\beta} = \sum_{n=1}^{\infty} \epsilon^n \sum_{m=0}^{n-1} A_{\mu\nu\alpha\beta}^{(m)} G^{(n-m)\alpha\beta} = \epsilon \, A_{\mu\nu\alpha\beta}^{(0)} G^{(1)\alpha\beta}  
% \nonumber \\&&\hspace{-.5cm} 
  + \epsilon^2 \left(A_{\mu\nu\alpha\beta}^{(0)} G^{(2)\alpha\beta}+ A_{\mu\nu\alpha\beta}^{(1)} G^{(1)\alpha\beta}\right) + 
 \ldots ,
\eea
and
\bea 
\label{expansion f G }
%&&\hspace{-.5cm} 
f(\Box) G^{\alpha\beta} = \sum_{n=1}^{\infty} \epsilon^n \sum_{m=0}^{n-1} f^{(m)} G^{(n-m)\alpha\beta} =   
%\\&&\hspace{-.5cm}
 \epsilon \, f^{(0)} G^{(1)\alpha\beta}   + \epsilon^2 \left(f^{(0)} G^{(2)\alpha\beta}+ f^{(1)} G^{(1)\alpha\beta}\right) + 
\ldots \, .
\eea
Let us consider the operator $Q_2({\rm Ric})$ and express it as
\bea
&&Q_2({\rm Ric}) = \sum_{n_1,n_2=0}^{\infty} \, c_{n_1,n_2}\,\sigma \, \left(D^{n_1} {\rm Ric}\right) \left(D^{n_2} {\rm Ric}\right) + \!\!\! \!\!\! 
\sum_{n_1,n_2,n_3=0}^{\infty} \, c_{n_1,n_2,n_3}\, \sigma^2 \, \left(D^{n_1} {\rm Ric}\right) \left(D^{n_2} {\rm Ric}\right)\left(D^{n_3} {\rm Ric}\right)+ \ldots \, ,\label{expansion Q2}
\eea
where ${\rm Ric}$ is the Ricci scalar or the Ricci curvature,  $D$ is a short notation for some  operator, e. g. $D= \sigma \Box$  or $D= g$,   $\, c_{n_1,n_2,n_3}$ are dimensionless parameters, and  the dots indicate a sum of terms of order higher than third in ${\rm Ric}$. Note that we have omitted indices in 
(\ref{expansion Q2}). We can  expand $D$ and ${\rm Ric}$ in powers of $\epsilon$ as
\be\label{expansion D}
D = \sum_{n=0}^\infty \epsilon^n D^{(n)} , \quad {\rm Ric} = \sum_{n=1}^\infty \epsilon^n {\rm Ric}^{(n)} \, ,
\ee
where we have used the fact that ${\rm Ric}^{(0)}(\eta)=0$.
From eq.(\ref{expansion Q2}) %it is easy to see that, if 
it follows that if 
${\rm Ric}^{(k)}= 0$ $\forall k < n$, then one has ${\rm Ric} = O(\epsilon^{n})$. Moreover, by means of  (\ref{expansion D}) one has $D = O(1)$, so that

\be \label{expansion Q}
{\rm Ric}^{(k)}= 0 \quad \forall \,\, k < n  \quad \Longrightarrow \quad Q_2({\rm Ric}) \sim \epsilon^{2n} \, .
\ee
Finally,  the Bianchi identity  can be expressed perturbatively by the means of (\ref{definition perturbation h and G}) and (\ref{definition perturbation nabla}) as
\bea 
\label{bianchi}
0=\nabla_\alpha G^{\alpha\beta} = \sum_{n=1}^{\infty} \epsilon^n \sum_{m=0}^{n-1} \nabla^{(m)}_\alpha G^{(n-m)\alpha\beta}  = \epsilon \, \partial_\alpha G^{(1)\alpha\beta} + \epsilon^2 \left(\partial_\alpha G^{(2)\alpha\beta}+ \nabla^{(1)}_\alpha G^{(1)\alpha\beta}\right) + O(\epsilon^3) \, .
\eea
Now we are ready to solve (\ref{einstein equation I}) perturbatively. At the lowest order $\epsilon^1$,  eq. (\ref{bianchi}) gives $\partial_\alpha G^{(1)\alpha\beta} = 0$, which also implies: 
\be\label{A0 G null}
A^{(0)}_{\mu\nu\alpha\beta} G^{(1)\alpha\beta} \propto \left(\eta_{\mu\nu} \partial_\alpha \partial_\beta - \eta_{\alpha\mu} \partial_\beta \partial_\nu\right) \gamma(\Box^{(0)}) G^{(1)\alpha\beta} = 0 \, ,
\ee
where we have used the fact that the operator $\gamma(\Box^{(0)})$ commute with the ordinary derivates  $\partial_\alpha$. Therefore, the operator in (\ref{expansion A G }) is null at order $\epsilon^1$. Moreover, since $G^{(0)}_{\mu\nu} = 0$ implies ${\rm Ric}^{(0)}= 0$ and ${\rm Ric}= O(\epsilon)$,  from eq.(\ref{expansion Q}) we see that the term $Q_2({\rm Ric})$ is at least of order $\epsilon ^2$, hence it does not contribute to the equations at the order $\ \epsilon^1$. In conclusion, in vacuum and at first perturbative order $\epsilon^1$, eq.(\ref{einstein equation I}) reads
\be \label{einstein equation order I}
f(\sigma \Box^{(0)}) G^{(1)}_{\mu\nu}=  \left(1 + \Box^{(0)} \, \gamma\left(\Box^{(0)} \right) \right) G^{(1)}_{\mu\nu} = 0 \, ,
\ee
and, since $f(\Box^{(0)})$ is invertible, this equations admits the unique solution $G^{(1)}_{\mu\nu} = 0$.

The invertibility of $f(\sigma \Box^{(0)})$ is  a simple consequence of the fact that  $f(z)$ is an entire analytic function without   zeros for finite complex $z$. In facts, under such hypothesis we can expand $f(\sigma \Box^{(0)})$ in power series of $\sigma$,  so that the Kernel of such operator will be given by the solutions of the following equation 

\be\label{expansion f phi}
f(\sigma \Box^{(0)}) \phi = \sum_{n=0}^{\infty} c_n \left(\Box^{(0)}\right)^n \sigma^n \, \phi = 0 \,
\ee
with $c_0\neq0$. Since (\ref{expansion f phi}) is analytic in $\sigma$, it  must vanish at any order, i.e.,  it must be $(\Box^{(0)})^n \, \phi = 0$, $\forall n \geq 0$, which implies that $\phi = 0$. Therefore, 
the operator $f(\sigma \Box^{(0)})$ is invertible because it is linear and its kernel is the zero function.
 An equivalent proof can be given using Fourier transforms, writing

\begin{equation}
f(\sigma \Box^{(0)}) \, \phi(x) = \int \frac{d^4k}{(2\pi)^4}\, f(-\sigma k^2) \, \tilde \phi(k) \, e^{i \, k\, x}\, ,
\end{equation}
where $\tilde{\phi}(k)$ is the Fourier transform of $\phi(x)$.
Since $f(z)$ has no zeros for finite $z$, one has $f(-\sigma k^2)\neq 0$, $\forall k^2 < \infty$, and the equation  $f(\sigma \Box^{(0)}) \, \phi(x) = 0$
has the only solution $\tilde \phi(k) = 0$, that is $\phi(x)= 0$.

Let us here emphasize the crucial role of the invertibility of the function $f(z)$ in our result. 
In order to better understand this point, let us consider a function $f(z)$ with a root of order $q$ in $z=m^2$, 
namely 
\be
f(\sigma \Box^{(0)}) G^{(1)}_{\mu\nu} = \left[ \sum_{n=0}^{\infty} c_n \left(\sigma\Box^{(0)} \right)^n \right] \left(\Box^{(0)} - m^2 \right)^q  \, G^{(1)}_{\mu\nu} = 0 \, ,
\ee
which has non null solutions $G^{(1)}_{\mu\nu}\neq 0$, duo to the solutions of the equation  
$\left(\Box^{(0)} - m^2\right)^q  \, G^{(1)}_{\mu\nu} = 0$. Therefore, if the function $f(\sigma \Box^{(0)})$ would be non invertible, we would have $G^{(1)}_{\mu\nu}\neq 0$, and our proof could not be implemented.

%This fact is intimately connected with another property of the nonlocal gravity (\ref{action gravity}) at quantum level: the fact that $f(z)$ has no zeros for finite complex values of $z$ implies that the graviton propagator has no poles other than the ones corresponding to the two polarizations of the graviton in Einstein theory, and the nonlocal gravity has no extra degrees of freedom, avoiding  ghosts or tachyons (aggiungere referenza). From the point of view of classical solutions, that means that,  if $f(z)$ had a zero of order  $q$ at some value of $z=m^2$, for $m^2<0$ the solutions  $h^{(1)}_{\mu\nu}$ would be  exponentially growing perturbations of the Minkowski metric, i. e., unstable modes propagating in the vacuum,  corresponding to ghost degrees of freedom in the quantum spectrum.

%
The outcome of our analysis of (\ref{einstein equation I}) at first  perturbative order in $\epsilon$ is that  $G^{(1)}_{\mu\nu}$ must vanish.  Before generalizing this result to any order, let us repeat our analysis at second order $\epsilon^2$. Using the Bianchi identity together with $G^{(1)}_{\mu\nu}=0$ we see that it must be $\partial_\alpha G^{(2) \alpha\beta} = 0$, which in turn gives
\be
\label{A0 G null2}
A^{(0)}_{\mu\nu\alpha\beta} G^{(2)\alpha\beta} \propto \left(\eta_{\mu\nu} \partial_\alpha \partial_\beta - \eta_{\alpha\mu} \partial_\beta \partial_\nu\right) \gamma(\Box^{(0)}) G^{(2)\alpha\beta} = 0 .
\ee
The above equation, together with $G^{(1)}_{\mu\nu}=0$, implies that $A_{\mu\nu\alpha\beta} G^{\alpha\beta}$ vanishes at second order in epsilon (see eq.(\ref{expansion A G })). Moreover,  $G^{(1)}_{\mu\nu}=0$ implies that ${\rm Ric}^{(1)}=0$, indeed from eq.(\ref{expansion Q}) it also comes  that $Q_2({\rm Ric})\sim \epsilon^4$.  Thus, at order $\epsilon^2$ eq. (\ref{einstein equation I}) reads
\be \label{einstein equation order II}
f(\sigma \Box^{(0)}) G^{(2)}_{\mu\nu}=  \left(1 + \Box^{(0)} \, \gamma\left(\Box^{(0)} \right) \right) G^{(2)}_{\mu\nu} = 0 \, ,
\ee
which, due to the  invertibility of  $f(\sigma \Box^{(0)})$, implies that $G^{(2)}_{\mu\nu}=0$.

Now it is easy to infer that eq.(\ref{einstein equation I}) implies that $G^{(n)}_{\mu\nu}=0$, $\forall n \geq 0$. This result can be proved by induction showing that, if $G^{(m)}_{\mu\nu}=0$, $\forall m \leq n$, then,  eq.(\ref{einstein equation I}) implies  
$G^{(n+1)}_{\mu\nu}=0$. %In facts, under such hypothesis,
Indeed, if $G^{(n)}_{\mu\nu}=0$  the Bianchi identity (\ref{bianchi}) gives $\partial_\alpha G^{(n+1) \alpha\beta}=0$, which  implies,  by the means of (\ref{expansion A G }), that $A_{\mu\nu\alpha\beta} G^{\alpha\beta} \sim \epsilon^{n+2}$. Moreover, since $G^{(m)}_{\mu\nu}=0$, $\forall m \leq n$ implies  ${\rm Ric}^{(m)} = 0$, $\forall m \leq n$, eq.(\ref{expansion Q}) tells us that $Q_2({\rm Ric}) \sim \epsilon^{2 (n+1)}$. Finally, using (\ref{expansion f G }) we conclude that at the order $\epsilon^{n+1}$, equation (\ref{einstein equation I}) 
 turns into $f(\sigma \Box^{(0)}) G^{(n+1)}_{\mu\nu}= 0$, which implies $G^{(n+1)}_{\mu\nu}=0$. 

Summarizing, we have proved that eq.(\ref{einstein equation I}) implies that, at any perturbative order in the gravitational perturbation, it must be $G^{(n)}_{\mu\nu}=0$. This, by means of eq.(\ref{definition perturbation h and G}), implies that the Einstein's tensor must vanish, namely

\be 
\label{G = 0}
G_{\mu\nu}(g_{\mu\nu}) =G_{\mu\nu}(\eta_{\mu\nu}  + \epsilon h_{\mu\nu})= \sum_{n=1}^\infty \epsilon^n G^{(n)}_{\mu\nu} = 0\, .
\ee
Therefore, the stability analysis for the Minkowski spacetime in NLG is exactly the same then in Einstein's gravity at all perturbative orders. 
Notice that the inverse implication is straightforward, since $G_{\mu\nu}=0$ implies that (\ref{einstein equation I}) is automatically satisfied.

Equation (\ref{G = 0})  makes evident that the stability of the Minkowski spacetime  is the same in NLG and in Einstein's gravity. Indeed,  the dynamics  of small perturbations around Minkowski is the same (at any perturbative order) in the two theories, which immediately allows us to infer about the evolution of small perturbations in NLG. Therefore, any Strongly Asymptotically Flat initial data set satisfying a Global Smallness Assumption leads to a unique smooth, geodesically complete, and asymptotically flat solution of eq.  (\ref{einstein equation I}) in the vacuum, which is in facts a solution of the Einstein's equations in the vacuum. This is a further confirmation of the result found in \cite{Dona:2015tra} where it was shown that all the $n$-points tree-level scattering amplitudes of NLG coincide with those in Einstein's gravity.

Let us discuss the validity of the perturbative approach used in the derivation of eq. (\ref{G = 0}).
This is based on the assumption that,  the metric and all the tensors constructed with it can be expanded in the quantity $|\epsilon \, h_{\mu\nu} | \ll 1$. Therefore we expand in powers of just one small parameter $\epsilon$, that represents the amplitude of the perturbations of the Minkowski metric, as expressed by eq. (\ref{definition h}).  This is a standard assumption----for instance it is the basis of all the post-Newtonian and post-Minkowskian studies of the gravitational  emission by compact objects, see \cite{blanchet}----and it is well justified by the choice of the class of small initial perturbations of Minkowski spacetime considered so far.

In facts, the GSA assumption implies that, at the initial time, the metric (\ref{definition h}) must satisfy the condition (\ref{GSA})  for a sufficiently small $\mu$, that implies that $\epsilon$ must be sufficiently small. Thus,  we can take $\epsilon$ small enough in order to guarantee that all the series expansions in $\epsilon$  are convergent at the  initial time.  
As a consequence, equation (\ref{G = 0}) is valid, and we infer that the evolution of $h_{\mu\nu}$ in NLG is the same as in general relativity. In \cite{Christodoulou2} it has been shown that, in the harmonic gauge and for asymptotically flat initial data satisfying the global smallness assumption, the solutions of vacuum Einstein equations converges asymptotically in time to Minkowski spacetime; and more precisely it has been shown that  $|h_{\mu\nu}| \lesssim t^{-1} \ln(t)$ converges asymptotically to zero. Using this result, we conclude that all the series expansions in $|\epsilon \, h_{\mu\nu}|$ are convergent at any time, since they are convergent at the initial time.  This proves the validity  of our perturbative scheme.

We emphasize that the condition of strong asymptotic flatness of the initial data is crucial for the time convergence to zero of the perturbation $h_{\mu\nu}$ proved in \cite{Christodoulou2}.  In facts, in \cite{briscese santini} it has been shown that, in the harmonic gauge, the collision of two plane gravitational waves produces a secular divergence of $h_{\mu\nu}$ and the break down of the series expansion (\ref{definition h}). However, the occurrence of this secularity is due to the fact that plane waves  do not satisfy the condition of asymptotic spatial flatness because they are not localized in the space but infinitely extended.
%, and they also carry infinite energy. L'ENERGIA DELL'ONDA PIANA E' FINITA
We also stress that the initial asymptotic flatness condition implies that the initial perturbation $h_{\mu\nu}(t_0,\vec{x})$ is confined into a finite volume $\mathcal{V}$ of the space, say $\vec{x} \in \mathcal{V}$. Thus,  far from that region $\mathcal{V}$, i.e. for $|\vec{x}|\rightarrow \infty$, one has $h_{\mu\nu}(t_0,\vec{x})\rightarrow 0$.
%, in such a way that the initial perturbation carries a finite energy. 
On the other hand, the asymptotic behavior $h_{\mu\nu} \sim t^{-1} \ln(t)$ for large times means that the perturbation $h_{\mu\nu}$, initially confined in $\mathcal{V}$, will be dinamically spread in all the space, to end up with the Minkowski metric. 
%, and its associated energy density converges to zero. 

Let us stress again the importance that $Q_2({\rm Ric})$ is at least quadratic in ${\rm Ric}$ in our derivation of the stability. To clarify this point, let us consider the following EOM:
\be\label{einstein equation modified}
f\left(\sigma \Box  \right) G_{\mu\nu}+\left(g_{\mu\nu} \nabla_\alpha \nabla_\beta - g_{\alpha\mu} \nabla_\beta \nabla_\nu\right)\gamma\left(\Box \right) G^{\alpha\beta}    = \sigma R_{\mu\tau\rho\sigma} R^{\,\,\,\tau\rho\sigma}_\nu + \dots\, , 
\ee
which is (\ref{einstein equation I}) with the replacement $Q_{2\mu\nu}  = - \sigma R_{\mu\tau\rho\sigma} R^{\tau\rho\sigma}_\nu + \dots$, so that $Q_{2\mu\nu}$ is now quadratic in the Riemann curvature. If we exploit the perturbative expansion of (\ref{einstein equation modified}) at the first order in $\epsilon$ we find
\be
G^{(1)}_{\mu\nu} = \sigma R^{(1)}_{\mu\tau\rho\sigma} R^{(0) \tau\rho\sigma}_\nu +\sigma R^{(0)}_{\mu\tau\rho\sigma} R^{(1) \tau\rho\sigma}_\nu + \dots \, 
\ee
which implies $G^{(1)}_{\mu\nu}=0$ and $R^{(1)}_{\mu\nu}=0$, but does not imply $R^{(1)}_{\alpha\beta\mu\nu}=0$. At the order  $\epsilon^2$ we get
\be
G^{(2)}_{\mu\nu} = \sigma R^{(2)}_{\mu\tau\rho\sigma} R^{(0) \tau\rho\sigma}_\nu 
+ \sigma R^{(0)}_{\mu\tau\rho\sigma} R^{(2) \tau\rho\sigma}_\nu 
+\sigma  R^{(1)}_{\mu\tau\rho\sigma} R^{(1) \tau\rho\sigma}_\nu+ \dots \, 
\ee
which does not imply $G^{(2)}_{\mu\nu} =0$ because of the term 
$R^{(1)}_{\mu\tau\rho\sigma} R^{(1) \tau\rho\sigma}_\nu \neq  0$. Therefore, if $Q_{2\mu\nu}$ in eq.(\ref{einstein equation I}) is not assumed at least quadratic in the Ricci curvature, with the exclusion of terms quadratic in the Riemann tensor, one does not get eq.(\ref{G = 0}), and can not conclude that small perturbations of the Minkowski metric are stable.

Also note that the outcome of this paper is in agreement with the unitarity of the theory (\ref{action gravity}) at quantum level \cite{brisceseUnitarity}. 
Once again, unitarity is guaranteed by the invertibility of the operator $f(\sigma \Box)$ in eq.(\ref{expansion f phi}), which is crucial for the derivation of eq.(\ref{G = 0}). Therefore, the stability of Minkowski spacetime at classical level is strongly related to the unitarity of the quantum theory.

Finally, we comment on the existence and occurrence of up to six extra degrees of freedom of the NLG (\ref{action gravity}), as it has been derived in  \cite{Calcagni modesto nardelli}. This theory  can be reformulated in terms of auxiliary fields, considering the following action

\be
S[g,\phi,\chi] = \frac{1}{2\kappa^2} \int d^4 x \sqrt{-g} \left[ R + 2 \, G_{\mu \nu}\,  \gamma(\Box) \, \phi^{\mu\nu} 
- \phi_{\mu\nu} \, \gamma(\Box) \, \phi^{\mu\nu} +R \, \gamma( \Box )\, \varphi
+ \frac{1}{2} \varphi \, \gamma(\Box) \, \varphi \, + \,V(\mathcal{R}) \right] ,\label{actionAUX}
\ee
containing the extra fields $\phi_{\mu\nu}$ and $\varphi$.
The EOM for the scalar $\varphi$ and the tensor $\phi_{\mu\nu}$ are  

\be\label{variation}
\frac{\delta S}{\delta \varphi} = 0 \quad \Longrightarrow \quad \varphi = G=- R \, , 
\qquad	
	\frac{\delta S}{\delta \phi^{\mu\nu} } = 0 \quad \Longrightarrow \quad \phi_{\mu\nu} = G_{\mu\nu} \,. 
	\ee
Notice that the auxiliary fields coincide with the Einstein's tensor and the Ricci scalar respectively. Furthermore, one has $\nabla^\mu \phi_{\mu\nu} =0$ and $\varphi=\phi^\mu_\mu$, so that one is left with up to six extra degrees of freedom.
Finally, eliminating the auxiliary fields from  (\ref{actionAUX}) we end up with  
(\ref{action gravity}), thus the two actions are equivalent.

It happens that the initial data satisfying the SAF condition and the GSA around the Minkowski metric set to zero the extra modes $\phi_{\mu\nu}$ because the evolution of such initial data obeys the Einstein's equations in vacuum (according to the equation (\ref{G = 0})), which set $\phi_{\mu\nu} = G_{\mu\nu} = 0$. On the other hand, the extra degrees of freedom could arise when considering more general initial data. However, such initial data will not be small perturbations of the Minkowski spacetime, because they imply that the corresponding spacetime has a curvature $R = O(\ell^{-2})$ at any time (including of course at the initial time), as we will show below. Since $\ell = \sqrt{\sigma}$ is the scale of nonlocality, which can be as small as the Planck length,  $R$ is huge, and a spacetime with such a large  curvature is not a globally small perturbation of the Minkowski spacetime. This explains why we do not see the modes $\phi_{\mu\nu}$ in our stability analysis. We stress that the  existence of classical exact solutions with $\phi_{\mu\nu} = G_{\mu\nu} \neq 0$ in the vacuum, makes evident that NLG is not identical to general relativity.

To explain these claims with more details, let us consider the classical exact solutions of the theory, and let us recast the exact equations of motion in the vacuum (\ref{einstein equation I}) as
\be \label{einstein equation I bis}
W_{\mu\nu\alpha\beta} \phi^{\alpha\beta}   =  - Q_{2\mu\nu}\left(\phi\right)  \, ,
\ee
where $\rm Ric$ has been  expressed in terms of $\phi$ using the relation $R_{\mu\nu}= G_{\alpha\beta}-g_{\alpha\beta}G/2= \phi_{\alpha\beta}-g_{\alpha\beta}\phi/2$, and the operator $W_{\mu\nu\alpha\beta}$ is defined as
\be \label{einstein equation I tris}
W_{\mu\nu\alpha\beta}\equiv\left(1 + \Box \, \gamma\left(\Box \right) \right) \delta_{\mu\alpha}\,\delta_{\nu\beta}+\left(g_{\mu\nu} \nabla_\alpha \nabla_\beta - g_{\alpha\mu} \nabla_\beta \nabla_\nu\right)\gamma\left(\Box \right) \, .
\ee
From the expression (\ref{expansion Q2}) of the operator $Q_2({\phi})$, 
one easily recognizes that equation (\ref{einstein equation I bis}) has the form
\be \label{einstein equation I quart}
W_{\mu\nu\alpha\beta} \phi^{\alpha\beta} = \ell^2 \, O(\phi^2) \,,
\ee
where, once more, $\ell$ is the length scale of nonlocality. 

%At that point, two situations can occur. 
Now two situations can occur: in the first case, which is the one considered in this paper, one has $\ell^2 \, |\phi| \ll 1$, so that the r.h.s. of (\ref{einstein equation I quart}) is negligible and one finds that $\phi_{\mu\nu}$ must be zero, as we concluded.   In other words, the only possibility compatible with the assumption that  the curvature does not exceed certain value, say  $|R| = |\phi|  \ll 1/\ell^2$, at the initial time,  is that curvature tensor must be zero at any time. Therefore, in that case, the equations of motion (\ref{einstein equation I bis}) coincide with the Einstein's equations in vacuum, i.e., $\phi_{\mu\nu} = 0$.

In the second case, in which $\ell^2 \, |\phi| \gtrsim 1$, the r.h.s. of (\ref{einstein equation I quart}) cannot be neglected, and one can have solutions of the equations of motion with $\phi_{\mu\nu}\neq 0$, so that the extra degrees of freedom can show up. However, in this case  $|R| = |\phi|  \gtrsim 1/\ell^2$, so that  the  curvature of the spacetime cannot be arbitrarily small, but it is  fixed by the scale of nonlocality. Since  $\ell$ is  very small, one understand that a spacetime with such a large curvature is not a globally small perturbation of the Minkowski spacetime.  Particularly, such a spacetime  cannot fulfil the SAF condition and the GSA around the Minkowski metric at the initial time.

Therefore,  choosing  the class of small initial perturbations of the Minkowski spacetime satisfying  the SAF condition and the GSA, one a  automatically sets to zero the extra degrees of freedom of the theory.
Such extra degrees of freedom   can be excited only on exact background solutions of  nonlocal gravity, that are not solutions of general relativity. In facts, if a spacetime is solution of Einstein's equations, it must contain only the graviton.  Since, any initial data satisfying the SAF condition and the GSA  evolves according to the vacuum Einstein's equations, such class of  initial data can excite only  the graviton. For the same reason,  background metrics containing  the extra degrees of freedom  cannot be seeded by small (in the SAF and GSA sense) perturbations of the Minkowski spacetime.

%Once more, we emphasize that,  choosing  the class of small (in the SAF and GSA sense) initial perturbations of the Minkowski spacetime, one a  automatically sets to zero the extra degrees of freedom of the theory.
%Such extra degrees of freedom   can be excited only on exact background solutions of  nonlocal gravity, that are not solutions of general relativity. In facts, if a spacetime is solution of Einstein's equations, it must contain only the graviton.  Since, any initial data satisfying the SAF condition and the GSA  evolves according to the vacuum Einstein's equations, such class of  initial data can excite only  the graviton.
%The extra degrees of freedom can be excited only on   exact background solutions of nonlocal gravity that cannot be seeded by small perturbations of the Minkowski spacetime.

Our result can be resumed as follows:  the NLG  (\ref{action gravity}) is stable under the same class of initial conditions under which GR is stable too, namely those satisfying the SAF condition and the GSA around Minkowski spacetime. Furthermore, such initial conditions do not excite the extra degrees of freedom
contained in the theory, and the graviton is the only propagating mode in the evolution of the initial data. The extra modes $\phi_{\mu\nu}$ can be nonzero in the vacuum, but their occurrence implies that the curvature of the spacetime must be very large $R=O(\ell^{-2})$ at any time, including  the initial time. This is why initial states containing the $\phi_{\mu\nu}$ modes are not small perturbations of the Minkowski spacetime. 

We stress that, since one is left with the two polarizations of general relativity,  the evolution of gravitational waves on a Minkowski background in NLG is the  same as in Einstein's gravity. Thus, NLG is in agreement with the recent observations of gravitational radiation from binary systems achieved in gravitational interferometers \cite{GW}.

The results presented in this paper can be extended to more general classes of metrics, including  Ricci flat and (A)dS spacetimes as will be proved elsewhere \cite{BCM}. So that, such spacetimes, e.g., (A)dS, will be stable in NLG provided they are stable in GR; for a review of (A)dS stability in general relativity see \cite{thai}.

\appendix\label{definition SAF GSA}

\section{Stability of the Minkowski spacetime in general relativity}
In this Appendix we recall the notions of Strongly Asymptotically Flat (SAF) condition and  Global Smallness Assumption (GSA) of  initial data sets, and  the stability theorem for the Minkowski spacetime in GR, as given in \cite{Christodoulou}. 

Let us first consider a foliation of the spacetime $\Sigma_\tau$ depending on a the time parameter $\tau$, such that $D\tau$ is always time-like. Indeed, the spacetime is diffeomorphic to a product manifold $\mathbb{R}\times \Sigma$, where $\Sigma$ is a three dimensional  manifold. Such a spacetime can be parameterized by the points of a slice $\Sigma_{\tau_0}$ by following the integral curves of $D\tau$, so that the metric in such reference frame takes the form
\be
ds^2 = \phi^2(\tau, \xi) \, d\tau^2 + \chi_{ij}(\tau, \xi)\, d\xi^i \, d\xi^j \, .
\ee
where $\chi_{ij}$ is a $3\times3$ Riemannian metric. Moreover, we define the following quantity, corresponding to the extrinsic curvature of the leaves $\Sigma_\tau$ as
\be 
\kappa_{ij} = \left(2 \phi\right)^{-1} \partial_\tau \chi_{ij} \, .
\ee 
An initial data set for the Einstein's equations is given by a set $\left(\Sigma_{\tau_0},\chi,\kappa\right)$ corresponding to initial conditions at some initial time $\tau_0$. We define the class of initial data satisfying the SAF condition as follows.

{\bf Definition} (SAF): we say that the initial data set  $\left(\Sigma_{\tau_0},\chi,\kappa\right)$ satisfies the SAF condition if there is a coordinate system $\left(\xi^1,\xi^2,\xi^3 \right)$ on $\Sigma_{\tau_0}$ such that, asymptotically for $|\xi|^2 = \sum_i \left(\xi^i \right)^2 \rightarrow \infty$ one has

\begin{eqnarray} 
\chi_{ij} = \left(1+ \frac{M}{\xi}\right)\delta_{ij} + o_4\left(\xi^{-3/2}\right)\, ,\nonumber\\
\kappa_{ij} =  o_3\left(\xi^{-5/2}\right)\, ,
\end{eqnarray}
where a function is said to be $o_m(\xi^{-n})$ if $\partial^l f\left(\xi\right) = o\left(r^{-n-l}\right)$ for 
$|\xi|^2  \rightarrow \infty$.

In order to express the GSA, we define the following quantity:
\begin{eqnarray} 
Q\left(\xi_0\right) = \sup_{\Sigma_{\tau_0}} \left\{\left(d_0^2+1\right)^3 \vert R^{(3)}_{ij}\vert^2 \right\} + \int_{\Sigma_{\tau_0}} \sum_{l=0}^{3} \left(d_0^2+1\right)^{l+1} \,\vert\nabla^{(3)\, l} \kappa\vert^2
+ \int_{\Sigma_{\tau_0}} \sum_{l=0}^{3} \left(d_0^2+1\right)^{l+3} \,|\nabla^l B_{ij}|^2 \, ,
\end{eqnarray}
where $d_0(\xi) = d(\xi_0,\xi)$ is the Riemannian geodesic distance between $\xi_0 \in \Sigma_{\tau_0}$ and $\xi \in \Sigma_{\tau_0}$. Moreover, the curvature $R^{(3)}_{ij}$ and the covariant derivatives  are constructed with the 3-metric $\chi_{ij}$, and $B_{ij}$ is the symmetric and traceless 2-tensor

\be 
B_{ij} = \epsilon_j^{ab} \nabla^{(3)}_a \left(R^{(3)}_{ib }  -\frac{1}{4}\chi_{ib}  R^{(3)}\right) \, .
\ee 

Now we can give the following definition (GSA): 
%{\bf Definition}: 
the metric $\chi_{ij}$ satisfies the GSA condition if there is a sufficiently small positive parameter $\mu$ such that 
\begin{eqnarray} \label{GSA}
\inf_{\xi_0\in\Sigma_{\tau_0}} Q\left(\xi_0\right) < \mu \, .
\end{eqnarray}

The stability of the Minkowski metric in general relativity GR is established by the following \cite{Christodoulou}.
{\bf Theorem}: any SAF initial data set which satisfies the GSA leads to a unique, globally hyperbolic, maximal, smooth, and geodesically complete solution of the Einstein Vacuum Equations foliated by a normal, maximal time foliation. Moreover, this development is globally asymptotically flat.

This theorem in facts means that it exists a $\mu_0$ such that, if the initial data satisfies the SAF condition and the GSA
with $\mu < \mu_0$, such initial data have a regular evolution (see for instance the discussion in \cite{Christodoulou2}). Therefore, the stability of Minkowski spacetime is inferred proving the existence of such $\mu_0$, without determining its size. Finally, note that the GSA (\ref{GSA}) for the metric (\ref{definition h})
($g_{\mu\nu}= \eta_{\mu\nu}+ \epsilon h_{\mu\nu}$) 
implies that $\epsilon$ must be sufficiently small. Therefore, the stability theorem  means that it exists a positive $\epsilon_0 < 1$ such that, if  $\epsilon < \epsilon_0$, the perturbations have a regular evolution.

\end{document}